\documentstyle[twocolumn,aps,epsf]{revtex}
                       \def\g{\gamma}        
                      \def\D{\Delta}        
           \def\s{\sigma}         \def\w{\omega}        
             \def\k{{\bf k}}               
\def\W{\Omega}
                  \def\Ek{E_{\bf k}}    
\def\S{{\cal S}}           \def\N{{\cal N}}             
\def\SA{{\cal S}_A}        \def\SF{{\cal S}_F}                          
\def\DS{{\cal D}_{\rm S}}  \def\DN{{\cal D}_{\rm N}}                    
                
\def\GFN{G_{\rm FN}}                                  
\def\RFN{R_{\rm FN}}       \def\RK{R_{\rm K}}                           
        \def\Peff{P_{\rm eff}}                       
\def\IA{I_{\rm A}}         \def\IF{I_{\rm F}}                           
\def\/{\over}              \def\<{\langle}         \def\>{\rangle}      
                          
\def\us{\uparrow}          \def\ds{\downarrow}     \def\dmu{\delta\mu}  
\def\[{\left[}             \def\]{\right]}                              
\def\({\left(}             \def\){\right)}                              
\def\tT{\tau_{\rm t}}      \def\tE{\tau_{\rm E}}   \def\tS{\tau_{\rm s}}

\def\aks{a_{\k\s}}         \def\gks{\g_{\k\s}}                          
\def\fks{f_{\k\s}}         \def\ks{\k\s}                                
\def\GS{{\tilde G}}        \def\GSus{{\tilde G}_\us}  \def\GSds{{\tilde G}_\ds}                 
\begin{document}
\twocolumn[\hsize\textwidth\columnwidth\hsize\csname
@twocolumnfalse\endcsname

%%%%%%%%%%%%%%%%%%%%%%%%%%%%%%%%%%%%%%%%%%%%%%%%%%%%%%%%%%%%%%%%%%%%
\title{    Spin-Imbalance and Magnetoresistance in                  
  Ferromagnet/Superconductor/Ferromagnet Double Tunnel Junctions}   
%%%%%%%%%%%%%%%%%%%%%%%%%%%%%%%%%%%%%%%%%%%%%%%%%%%%%%%%%%%%%%%%%%%%

\author{S. Takahashi$^1$, H. Imamura$^2$, and S. Maekawa$^1$ }

\address{$^1$Institute for Materials Research, Tohoku University, Sendai
980-8577, Japan}
\address{$^2$CREST and Institute for Materials Research, Tohoku University,
Sendai 980-8577, Japan}
%-----------------------------------------------------------------------------
\date{January 22, 1999}

\maketitle

%%%%%%%%%%%
% Abstract 
%%%%%%%%%%%
{
\begin{abstract}
We theoretically study the spin-dependent transport in a
ferromagnet/super- conductor/ferromagnet double tunnel junction.
The tunneling current in the antiferromagnetic alignment of the
magnetizations gives rise to a spin imbalance in the superconductor.
The resulting nonequilibrium spin density strongly suppresses the
superconductivity with increase of bias voltage and destroys it at a
critical voltage $V_c$.   The results provide
a new method not only for measuring the spin polarization
of ferromagnets but also for controlling superconductivity and
tunnel magnetoresistance (TMR) by applying the bias voltage.
\end{abstract} 
}
\vskip0.3pc]

Since the early experiments demonstrated the spin-polarized tunneling of
electrons from ferromagnetic metals (FM) into superconductors (SC) in FM/SC
junctions \cite{meservey}, the concept of the spin-polarized transport
has been of vital importance in magnetic junctions and multilayers.
Firstly the tunneling currents strongly depends on the relative orientation
of magnetizations in FM/FM tunnel junctions; the tunnel resistance
decreases when the magnetizations are aligned in a magnetic field,
causing tunnel magnetoresistance (TMR) \cite{jullie,mae,mood,miya}.  
Secondly the spin-polarized current driven from a FM into a normal metal
(N) or superconductor (SC) gives rise to a nonequilibrium spin density
in N or SC \cite{johnson,aronov,johnsonS}.
In a FM/N/FM double junction the TMR effect is brought about by
accumulation of spin-polarized electrons in N 
 \cite{fnf}.  %%% \cite{brataas,barnasN,imamura}.
In a FM/SC/FM double junction \cite{johnsonS}, on the other hand, we
expect the strong competition between superconductivity and magnetism
induced by the spin polarization in SC.   Of particular interest in
the FM/SC/FM structure is not only to find novel magnetoresistive
effects due to the competition but also application to
magnetoelectronics.

In this Letter we show that a FM/SC/FM double tunnel junction is a new
magnetoresistive device to control superconductivity by applying the bias
voltage (or current).
In the antiferromagnetic (A) alignment of magnetizations (see Fig.~1),
a nonequilibrium spin density is induced in SC due to the imbalance in
the tunneling currents carried by the spin-up and spin-down electrons,
so that the superconducting gap $\D$ is reduced with increasing
bias voltage and vanishes at a critical voltage $V_c$.  In the
ferromagnetic (F) alignment, however, there is no spin-density in SC.
Consequently, TMR has a strong voltage dependence around $V_c$;
TMR is enhanced compared with that in the normal state above $V_c$,
while it changes sign to show an inverse TMR effect for
some voltage range below $V_c$.
It is shown that $V_c$ is inversely proportional to the spin polarization
$P$ of FM ($V_c \propto 1/P$), which provides a new method for determining
$P$ of FM.

We consider a FM/SC/FM double tunnel junction as shown in Fig.~1.
The left and right electrodes are made of the same FM and the central
one is a thin film SC.  The magnetization of the left FM is chosen to
point up and that of the right FM is either up or down.
The voltages $-V_1$ and $V_2$ ($=V-V_1$) are applied to the left and
right electrodes, respectively.   We assume that the energy relaxation
time $\tE$ of quasiparticles in SC is shorter than the time
$\tT$ between two successive tunneling events, 
whereas the spin relaxation time $\tS$ is longer than $\tT$.
Consequently, the electrons tunneling into SC relax to              \topskip 0cm
the Fermi distribution before leaving SC, keeping their 
spin direction during the stay in SC.

%=================== FIG. 1 ===================
\begin{figure}
    \epsfxsize=0.8\columnwidth
\centerline{\hbox{
      \epsffile{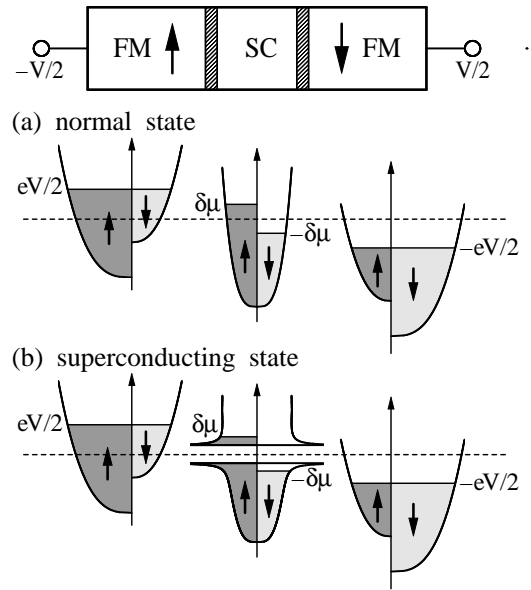}}
      }
\caption{ Double tunnel junction consisting of two ferromagnets
(FM) and a superconductor (SC) separated by insulating barriers
in the case of identical barriers.
Schematic densities of states of FMs (left and light) and SC (middle)
in the antiparallel alignment of magnetizations in FMs are shown when SC
is in the normal state (a) and in the superconducting state (b).
$\dmu$ denotes the shift of the chemical potentials in SC.
%%% for the spin subbands in SC.
    }
\end{figure}

We calculate the tunneling current using a phenomenological tunneling
Hamiltonian.   If SC is in the superconducting state, it is convenient
to rewrite the electron operators $a_{\k\s}$ in SC in terms of
quasiparticle operators $\g_{\k\s}$ appropriate to the superconducting
states, using the Bogoliubov transformation,
$a_{\k\us} = u_\k\g_{\k\us} + v_\k^*\g^\dagger_{-\k\ds}$ and
$a_{\k\ds} = u_{-\k}\g_{\k\ds} - v_{-\k}^*\g^\dagger_{-\k\us}$,
where $|u_\k|^2 = {1\/2}\( 1+{\xi_\k/E_\k}\)$,
      $|v_\k|^2 = {1\/2}\( 1-{\xi_\k/E_\k}\)$, and
$E_\k=[{\xi_\k^2+\D^2}]^{1/2}$ is the quasiparticle dispersion of
SC, $\xi_\k$ being the one-electron energy relative to
the chemical potential and $\D$ being the gap parameter.
Then, using the golden rule formula, we calculate the spin-dependent
currents $I_{j\s}$ across the $j$th junction.  The results are
  %----------------------------------(1)------------------------
\begin{mathletters}                                             
  \begin{eqnarray}                                              
    I_{1\us}  &=&  \({G_{1\us}/e}\) \[ \N_1 - \S - Q^*/2 \] ,   
            \label{eq:I1-u} \\                                  
    I_{1\ds}  &=&  \({G_{1\ds}/e}\) \[ \N_1 + \S - Q^*/2 \] ,   
            \label{eq:I1-d} \\                                  
    I_{2\us}  &=&  \({G_{2\us}/e}\) \[ \N_2 + \S + Q^*/2 \] ,   
            \label{eq:I2-u} \\                                  
    I_{2\ds}  &=&  \({G_{2\ds}/e}\) \[ \N_2 - \S + Q^*/2 \] ,   
            \label{eq:I2-d}                                     
  \end{eqnarray}                                                
\end{mathletters}                                               
  %-------------------------------------------------------------
\noindent
where $G_{j\s}$ is the tunnel conductance of the  $j$th
junction for electrons with spin $\s$ if SC is in the normal state.
The quantity $\N_j$ is given by the usual expression  \cite{tinkham}
  %---------------------(2)---------------------
  \begin{equation}                              
    \N_j  =  \int_{\D}^\infty  \DS(\Ek)         
    \Bigl[ f_0\bigl(\Ek-{eV_j}\bigr)            
- f_0\bigl(\Ek+{eV_j}\bigr)\Bigr] d\Ek  ,       
     \label{eq:N}                               
  \end{equation}                                
  %---------------------------------------------
where $\DS(\Ek)=\Ek/\sqrt{\Ek^2-\D^2}$ is the normalized BCS density of
states and $f_0(\Ek\pm{eV_j})$ is the Fermi distribution function of
thermal equilibrium in FM.
The quantity $\S$  %%%in Eqs.~(\ref{eq:I1-u}) - (\ref{eq:I2-d})
represents the spin density normalized to the density of states
of SC in the normal state $\DN$ and is given by
  %----------------------(3)--------------------
  \begin{equation}                              
     \S  =  \int_{\D}^\infty  \DS(\Ek)          
         \( f_{\k\us} - f_{\k\ds}  \) d\Ek ,    
     \label{eq:S}                               
  \end{equation}                                
  %---------------------------------------------
where $f_{\k\s}= \<\g^\dagger_{\k\s} \g_{\k\s}\>$ is the distribution function
of quasiparticles with energy $E_\k$ and spin $\s$ in SC.  The quantity
$Q^*$ is the charge density normalized to $e\DN$ due to the imbalance in
populations of electronlike and holelike quasiparticles \cite{tinkham}
  %--------------------(4)----------------------
  \begin{equation}                              
     Q^*  =  2 \sum_\s \int_{\D}^\infty         
         \( f_{\k\s}^> - f_{\k\s}^< \) d\Ek ,   
     \label{eq:Q}                               
  \end{equation}                                
  %---------------------------------------------
where $f_{\k\s}^>$ and $f_{\k\s}^<$ represent $f_{\k\s}$ in the
electronlike ($k>k_F$) and holelike ($k<k_F$) branches of $E_\k$,
respectively.

In the limit of vanishing spin-flip scattering, the spin-up and spin-down
currents are treated as independent channels.  The conservation of
the currents  at junctions 1 and 2, $I_{1\s} = I_{2\s}$, yields the relations
  %-----------------------------------(5,6)-----------------------------
  \begin{eqnarray}                                                      
   \S  &=& \[\( G_{1\us}G_{2\ds} - G_{1\ds}G_{2\us} \)/ (\GSus\GSds) \] 
           \(\N_1 + \N_2\)/2,                                           
     \label{eq:SS}                          \\                          
       &Q&^* = \sum_\s \[ ( {G_{1\s}/\GS_\s} ) {\N_1}                   
         -          ( {G_{2\s}/\GS_\s} ) {\N_2} \],                     
     \label{eq:QQ}                                                      
  \end{eqnarray}                                                        
  %---------------------------------------------------------------------
where ${\tilde G}_\s = G_{1\s} + G_{2\s}$.
If the tunnel barriers are symmetric [see Fig. 1],
$V_{1}=V_{2} = V/2$ and $\N_{1}=\N_{2} = \N$,
so that $Q^*=0$ for both alignments.
This is because the charge transport is symmetric at the two junctions
where the injected and extracted charges are balanced. %%% \cite{heslinga}.
In the following we assume the identical tunnel barriers and
neglect the effect of the charge imbalance.

The spin density $\S$ depends strongly on whether the magnetizations
in FMs are parallel or antiparallel.
The spin density $\SA$ in the {A-alignment} ($G_{1\s} = G_{2{-\s}}$)
satisfies the relation
  %------------------(7)-----------------
  \begin{equation}                       
      \SA = P \N ,   \label{eq:S=PN}     
  \end{equation}                         
  %--------------------------------------
where $P=|G_{j\us}-G_{j\ds}|/\(G_{j\us}+G_{j\ds}\)$ represents
the spin polarization of FM \cite{meservey}.
The relation (\ref{eq:S=PN}) implies that a finite $\S_A$ is induced by
applying the bias voltage.   The $\SA \ne 0$ is a consequence
of symmetry breaking of the spin transport in the A-alignment, and
is realized when the distribution of the spin-up and spin-down
quasiparticles is disequilibrium, i.e., $f_{\k\us} \ne f_{\k\ds}$,
as seen from Eq.~(\ref{eq:S}).  However, the spin density $\S_F$ in the
{F-alignment} ($G_{1\s}=G_{2\s}$) has no net spin-density in SC ($\SF = 0$)
% %------------------(8)-----------------
% \begin{equation}                       
%     \SF = 0  ,    \label{eq:S=0}      
% \end{equation}                         
% %--------------------------------------
due to the symmetric spin transport at the junctions.
 %%% It is worthwhile to note that the relations (\ref{eq:S=PN}) and
 %%% (\ref{eq:S=0}) are also hold for the case of asymmetric
 %%% tunnel barriers.  ($|{\hat T}_1|^2 \ne |{\hat T}_2|^2$).
 %%%, as long as the same FM are used for the electrodes.

The distribution function $f_{\k\s}$ is determined as follows.
When the thickness of SC is much smaller than the spin diffusion length
 \cite{johnson}, the distribution of spin-up and spin-down quasiparticles
is spatially uniform in SC.   For $\tE < \tT < \tS$, the distribution
is described by the Fermi function $f_0$, but the chemical potentials  %%$\mu_\us$ and $\mu_\ds$
of the spin-up and spin-down quasiparticles are
shifted oppositely by $\dmu$ from the equilibrium one (see Fig.~1) to produce
the nonequilibrium spin density.  Thus we write $f_{\k\s}$ as \cite{spin}
  %------------------(9)----------------
  \begin{equation}                      
      f_{\k\us} = f_0(\Ek - \dmu), \ \ \ \
      f_{\k\ds} = f_0(\Ek + \dmu).      
     \label{eq:fks}                     
  \end{equation}                        
  %-------------------------------------
In the normal state ($\D=0$), from Eqs.~(\ref{eq:S}), (\ref{eq:S=PN}),
and (\ref{eq:fks}), we have $\dmu_A=\SA={1\/2} PeV$
in the A-alignment 
\cite{fnf},  %%% \cite{brataas,barnasN,imamura},
whereas $\dmu_F=\SF=0$ in the F-alignment.

We first discuss how the superconductivity is affected by the
nonequilibrium spin imbalance in SC.
The gap $\D$ in the nonequilibrium situation is determined by $\fks$
through the BCS gap equation \cite{tinkham}
  %-----------------------(10)---------------------
  \begin{eqnarray}                                
    {1\/\DN V_{\rm BCS}} = \int_0^{\hbar\w_D} d\xi_\k 
      { 1 - f_{\k\us} - f_{\k\ds}\/ E_\k } ,
    \label{eq:gap}                                
  \end{eqnarray}                                  
  %-----------------------------------------------
where $f_{\ks}$ is given by Eq.~(\ref{eq:fks}).
We note that Eq.~(\ref{eq:gap}) with Eq.~(\ref{eq:fks}) is %%% exactly
the same as that of SC in the paramagnetic limit if $\dmu$ is taken to be
the Zeeman energy $\mu_{\rm B}H$ \cite{sarma}.   The chemical potential
difference $2\dmu$ plays the role of pair breaking energy. Therefore
the superconductivity is destroyed in the A-alignment when $\dmu$
exceeds a certain critical value by increasing the voltage.
To show this, we solve self-consistently Eqs.~(\ref{eq:S=PN}) and
(\ref{eq:gap}) with respect to $\D$ and $\dmu$, and obtain $\D$ and $\S$
as functions of $V$.

Figure 2(a) shows the gap parameter $\D_A$ in the A-alignment as a
function of bias voltage $V$ for $P=0.4$, the spin polarization of
Fe \cite{meservey}.
The quantity $\D_{0}$ denotes the value of $\D_A$ for $P=0$ at $T=0$.
The gap parameter $\D_A$ decreases with increasing $V$ and vanishes
at the critical voltage $V_c$.
 %%% At $T=0$, the critical voltage is given by $eV_c/2\D_0=1/2P$.
At very low temperatures $\D_A$ becomes multi-valued in a certain
range of $eV$ just below $2\D_{0}$; 
At $T=0$ it has three solutions, $\D_A=\D_{0}$ and
$\D_A$ = $\D_{0} [1-2P^2\pm 2P\sqrt{(eV/2\D_{0})^2+P^2-1}]^{1/2}$,
in the range $0.92 < eV/2\D_{0} < 1$.
When $V$ is increased (or decreased) at $T \sim 0$,
an instability into a spatially inhomogeneous state with different $\D_A$
takes place at a certain voltage within the range.
At $V_c$ where $\D_A = 0$ and $\dmu_A = {1\/2}PeV_c$,
Eq.~(\ref{eq:gap}) reduces to an implicit equation for $PeV_c$ and $T$,
%% \begin{equation}
%%  \ln{T\/T_c} + \psi\({1/2}+{PeV\/2\pi T}\)-\psi\({1\/2}\) =0 .
%% \label{eq:VcT}
%% \end{equation}
which gives a universal relation between $PeV_c/\D_{0}$ and $T/T_c$,
as shown in the inset of Fig.~2(a).
In particular, we have $eV_c=\D_0/P$ at $T=0$.   Therefore, we can determine
the spin polarization of FM by measuring $V_c$.
Since the paramagnetic effect caused by spin accumulation becomes
stronger with decreasing $T$, $V_c$ is not a monotonic function of
$T$, but has a maximum at $T/T_c=0.5$.
Figure 2(b) shows the voltage dependence of the spin density $\SA$
in the A-alignment.   The dotted line indicates the values of
$\SA={1\/2}PeV$ in the normal state.    As $T$ is lowered below $T_c$,
$\SA$ is suppressed below $V_c$ by the opening of the energy gap.
At and near $T=0$, $\SA$ shows the S-shaped anomaly around $eV_c \sim 2\D_{0}$,
which stems from the multiplicity of $\D_A$ shown in Fig.~2(a).
The detailed behavior of the anomaly is shown in the inset of Fig.~2(b).
In the F-alignment, $\D_F$ has no $V$ dependence and has the same value as $\D_A$($V=0)$.

We now calculate the tunneling current as a function
of bias voltage $V$.  From Eqs.~(2a)-(2d), the total currents
%=================== FIG. 2 ===================
\begin{figure}
    \epsfxsize=0.8\columnwidth
\centerline{\hbox{
      \epsffile{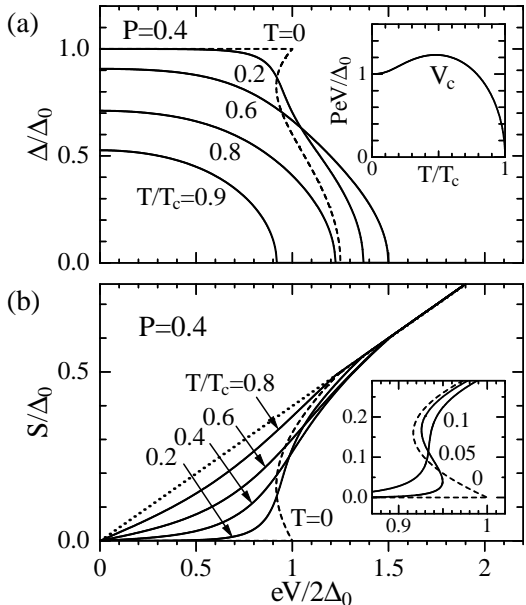}}}
\caption{
(a) Gap parameter $\D_A$ in SC as a function of bias voltage $V$ for
different temperatures below $T_c$ in the antiferromagnetic alignment.
The inset shows the critical voltage $V_c$ vs temperature $T$.
(b) Spin density $\S_A$ in SC as a function of $V$.
The inset shows an enlarged view of $\S_A$ around $eV/2\D_0=1$ at $T/T_c=$ 0,
0.05, and 0.1.
    }
\end{figure}
\noindent$\IF$ and $\IA$ in the F and A alignments are given by
  %---------------------------------------------
  \begin{eqnarray}                              
     \IF(V)  &=&  \(\GFN / e\) \N(V,\D_F),      
     \label{eq:IF}      \\                      
     \IA(V)  &=&  \(\GFN / e\) \(1-P^2\) \N(V,\D_A),
     \label{eq:IA}                              
  \end{eqnarray}                                
  %---------------------------------------------
where $\N(V,\D)$ is given in Eq.~(\ref{eq:N}) and
$\GFN=G_{j\us}+G_{j\ds}$.  %%% the junction conductance.
It follows from Eqs.~(\ref{eq:S=PN})
 and (\ref{eq:IA}) that $\IA \propto \SA$,
so that Fig.~2(b) also represent 
the $V$ dependence of $\IA$.

Figure 3(a) shows the voltage dependence of the differential conductance
$G_F$ and $G_A$ for the F and A alignments at $T/T_c=0.4$.  The $G_F$ shows
the ordinary dependence on $V$ expected for the constant gap $\D_F$.
In contrast, because of the decrease in $\D_A$ with increasing voltage,
$G_A$ increases with voltage more rapidly than $G_F$, forming a higher
peak than $G_F$, and then decreases steeply.
At $V_c$,
$G_A$ jumps to the conductance $G_A^N$ in the normal state. 
The tunnel magnetoresistance (TMR) is calculated by the formula: TMR 
$= \(G_F/G_A\)-1$.  Using the values of Fig.~3(a), we obtain the
$V$ dependence of TMR shown in Fig.~3(b).   
At $V=0$ where $\D_A = \D_F$, TMR takes
the same value as in the normal state.  A deep negative dip appears
at $eV/2\D_{0} \sim 1$ where $\D_A$ steeply decreases, %%%% [see Fig.~3(a)],
exhibiting an inverse TMR effect ($G_A>G_F$), and is followed by
the discontinuous jump at $V_c$ above which TMR is highly
enhanced compared to that in the normal state.

The relation $I_A \propto \S_A$ in the A-alignment directly indicates that
the superconductivity of SC is strongly suppressed  with increase of
injection current $I_A$.  Using the 
relation and the result of $\D_A$ and $\S_A$ in Fig.~2, we obtain 
$\D_A$ as a function of injection
current $I_A$.  Figure 4 shows 
%=================== FIG. 3 ===================
\begin{figure}
    \epsfxsize=0.8\columnwidth
\centerline{\hbox{
      \epsffile{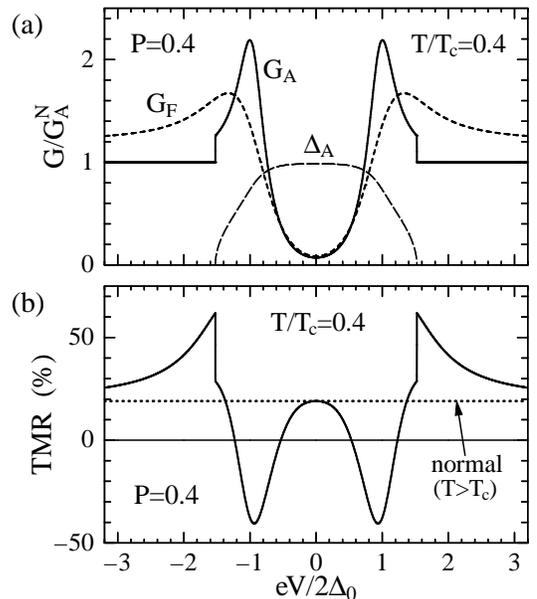}}
      }
\caption{
(a) Tunnel conductance as a function of bias voltage.
The dashed and solid curves indicate the conductances $G_F$ and $G_A$
for the ferromagnetic and antiferromagnetic alignments, respectively.
(b) Tunnel magnetoresistance (TMR) as a function of bias voltage.
The dotted line indicates TMR $= P^2/(1-P^2)$ in the normal state.
  }
\end{figure}

%==================== FIG. 4 ==================
\begin{figure}
    \epsfxsize=0.9\columnwidth
\centerline{\hbox{
      \epsffile{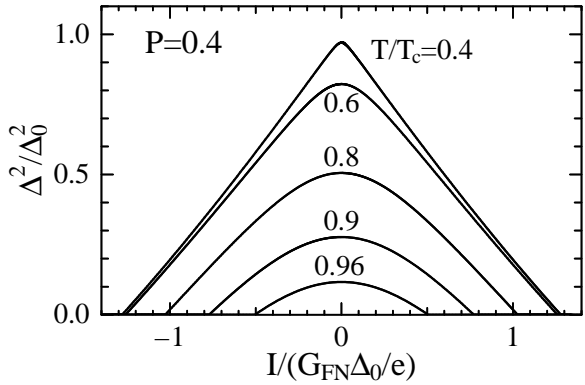}}
      }
\caption{
Square of $\D_A$ vs injection current $I_A$ at different temperatures
below $T_c$.   $\D^2_A$ is proportional to the superfluid density
and thus represents the critical current of SC.
    }
\end{figure}

\noindent
the square of $\D_A$ as a function of $I_A$.
Since $\D_A^2$ is proportional to the superfluid density which is a
measure of the critical current $I_c$ of SC,  Fig.~4 represents also
the critical current as a function of the spin-injection current.

The critical current suppression by spin injection has been observed
in FM/SC heterostructures made of a high-$T_c$ SC and a ferromagnetic
manganite with $P\sim 100 \%$ \cite{vasko,dong}.  The experimental
result of $I_c$ is surprisingly similar to that shown in Fig.~4. 
This strongly suggests that the injection currents from FM build
up the spin density in SC of the heterostructures.   Since the
spin density is accumulated more efficiently in the double-junction
geometry, the FM/SC/FM double junctions using high-$T_c$ SCs and
ferromagnetic manganites is quite promising to test our predictions.

Another candidate for SC in the FM/SC/FM junction is a thin film of
clean SC with sufficiently long spin-relaxation time such as Al.
The depression of the TMR effect is caused predominantly by spin
relaxation due to spin-orbit scattering in SC.
Here, we derive a condition for observing the TMR effect when SC is
in the normal state.   By balancing the population rate
$(I_{1\us}-I_{2\us})/e$ with the relaxation rate $Ad\DN\S_A/\tS$,  %%% \cite{fnf},
where $d$ is the thickness of SC and $A$ the junction area, we obtain
$\S_A={1\/2}\Peff eV$ with the effective spin polarization
$\Peff=P/\(1+\tT/\tS\)$ where $\tT = Ad\DN h(\RFN/\RK)$
 \cite{heslinga}
with $\RK=h/e^2\approx 26$ k$\Omega$ and $\RFN=1/\GFN$.  To retain a
substantial value of $\Peff$, we are required to satisfy $\tT \alt \tS$.
For the case of clean Al with the values of $\tS \sim 10^{-8}$ s
 \cite{johnson}
and $\DN$ $\sim 10^{22}$/(eV-cm$^3$), we have the condition
$\RFN A \alt (63/d)\times 10^{-5}$ $\Omega$ cm$^2$
for the specific junction resistance $\RFN A$ and $d$ (\AA).

In the above calculations we have not considered the effect of Andreev
reflection (AR) on the tunnel conductance, since the resistance of a
tunnel junction with a thin insulating layer is much higher than that
of a metallic contact, i.e.,
$\RFN A \gg \pi\RK/2k_F^2 \approx 10^{-11}$ $\W$ cm$^2$,
where $k_F$ is the Fermi momentum.
In metallic contacts with resistance comparable to $\pi\RK/2k_F^2$,
on the other hand, the conductance is dominated by AR for $eV < 2\D$
  \cite{blonder}.
Recently, the suppression of AR in FM/SC nanocontacts has been
used to measure $P$ in ferromagnets
  \cite{soulen}.

In summary, we have studied the spin-dependent tunneling in a FM/SC/FM
double tunnel junction.   The spin imbalance in the tunneling currents
gives rise to the nonequilibrium spin density
in SC.  The superconductivity is strongly suppressed with increase
of bias voltage and destroyed at the critical voltage $V_c$ $(\propto 1/P)$.
The tunnel magnetoresistance exhibits an unusual voltage dependence
around $V_c$ below $T_c$.   
The results predicted in this Letter provide a method for measuring
the spin polarization of FMs as well as for controlling
superconductivity and TMR by application of bias voltage.

We are grateful to J.S.~Moodera for valuable discussions for experimental
set-up.  This work is supported by a Grant-in-Aid for Scientific Research
Priority Area for Ministry of Education, Science and Culture of Japan
and NEDO Japan.

%%%%%%%%%%%%%%%%%%%%%%%%%%%
%        References        
%%%%%%%%%%%%%%%%%%%%%%%%%%%


\begin{references}
\bibitem{spin}
The spin splitting in the normal state is described by the term 
$-\sum_{\k\s}\s\dmu\aks^\dagger\aks$, which is transformed into
$-\sum_{\k\s}\s\dmu\gks^\dagger\gks$ by the Bogoliuvov transformation.
Thus the quasiparticle excitation energy is given by $E_{\k}-\s\dmu$
in the superconducting state.
\end{references}
\end{document}